\documentclass{article}
\usepackage{amsmath, amssymb, amsfonts}
\title{THE MILNE SPACETIME AND THE HADRONIC RINDLER HORIZON } 
\author{Hristu Culetu, \\Ovidius University, Dept.of Physics, \\B-dul Mamaia 124, 900527 Constanta, Romania, \\e-mail : hculetu@yahoo.com}

\begin{document}
\numberwithin{equation}{section}
\pagenumbering{arabic}
\maketitle
\newcommand{\fv}{\boldsymbol{f}}
\newcommand{\tv}{\boldsymbol{t}}
\newcommand{\gv}{\boldsymbol{g}}
\newcommand{\OV}{\boldsymbol{O}}
\newcommand{\wv}{\boldsymbol{w}}
\newcommand{\WV}{\boldsymbol{W}}
\newcommand{\NV}{\boldsymbol{N}}
\newcommand{\hv}{\boldsymbol{h}}
\newcommand{\yv}{\boldsymbol{y}}
\newcommand{\RE}{\textrm{Re}}
\newcommand{\IM}{\textrm{Im}}
\newcommand{\rot}{\textrm{rot}}
\newcommand{\dv}{\boldsymbol{d}}
\newcommand{\grad}{\textrm{grad}}
\newcommand{\Tr}{\textrm{Tr}}
\newcommand{\ua}{\uparrow}
\newcommand{\da}{\downarrow}
\newcommand{\ct}{\textrm{const}}
\newcommand{\xv}{\boldsymbol{x}}
\newcommand{\mv}{\boldsymbol{m}}
\newcommand{\rv}{\boldsymbol{r}}
\newcommand{\kv}{\boldsymbol{k}}
\newcommand{\VE}{\boldsymbol{V}}
\newcommand{\sv}{\boldsymbol{s}}
\newcommand{\RV}{\boldsymbol{R}}
\newcommand{\pv}{\boldsymbol{p}}
\newcommand{\PV}{\boldsymbol{P}}
\newcommand{\EV}{\boldsymbol{E}}
\newcommand{\DV}{\boldsymbol{D}}
\newcommand{\BV}{\boldsymbol{B}}
\newcommand{\HV}{\boldsymbol{H}}
\newcommand{\MV}{\boldsymbol{M}}
\newcommand{\be}{\begin{equation}}
\newcommand{\ee}{\end{equation}}
\newcommand{\ba}{\begin{eqnarray}}
\newcommand{\ea}{\end{eqnarray}}
\newcommand{\bq}{\begin{eqnarray*}}
\newcommand{\eq}{\end{eqnarray*}}
\newcommand{\pa}{\partial}
\newcommand{\f}{\frac}
\newcommand{\FV}{\boldsymbol{F}}
\newcommand{\ve}{\boldsymbol{v}}
\newcommand{\AV}{\boldsymbol{A}}
\newcommand{\jv}{\boldsymbol{j}}
\newcommand{\LV}{\boldsymbol{L}}
\newcommand{\SV}{\boldsymbol{S}}
\newcommand{\av}{\boldsymbol{a}}
\newcommand{\qv}{\boldsymbol{q}}
\newcommand{\QV}{\boldsymbol{Q}}
\newcommand{\ev}{\boldsymbol{e}}
\newcommand{\uv}{\boldsymbol{u}}
\newcommand{\KV}{\boldsymbol{K}}
\newcommand{\ro}{\boldsymbol{\rho}}
\newcommand{\si}{\boldsymbol{\sigma}}
\newcommand{\thv}{\boldsymbol{\theta}}
\newcommand{\bv}{\boldsymbol{b}}
\newcommand{\JV}{\boldsymbol{J}}
\newcommand{\nv}{\boldsymbol{n}}
\newcommand{\lv}{\boldsymbol{l}}
\newcommand{\om}{\boldsymbol{\omega}}
\newcommand{\Om}{\boldsymbol{\Omega}}
\newcommand{\Piv}{\boldsymbol{\Pi}}
\newcommand{\UV}{\boldsymbol{U}}
\newcommand{\iv}{\boldsymbol{i}}
\newcommand{\nuv}{\boldsymbol{\nu}}
\newcommand{\muv}{\boldsymbol{\mu}}
\newcommand{\lm}{\boldsymbol{\lambda}}
\newcommand{\Lm}{\boldsymbol{\Lambda}}
\newcommand{\opsi}{\overline{\psi}}
\renewcommand{\tan}{\textrm{tg}}
\renewcommand{\cot}{\textrm{ctg}}
\renewcommand{\sinh}{\textrm{sh}}
\renewcommand{\cosh}{\textrm{ch}}
\renewcommand{\tanh}{\textrm{th}}
\renewcommand{\coth}{\textrm{cth}}

\begin{abstract}
 A direct relation between the time dependent Milne geometry and the Rindler spacetime is shown. Milne's metric corresponds to the region beyond Rindler's event horizon (in the wedge $t \succ |x|)$). We point out that inside a Schwarzschild black hole and near its horizon, the metric may be Milne's flat metric.
 
 It was found that the shear tensor associated to a congruence of fluid particles of the RHIC expanding fireball has the same structure as that corresponding to the anisotropic fluid from the black hole interior, even though the latter geometry is curved.\\

 \textbf{Keywords} :  Near horizon metric; Rindler horizon; Shear viscosity.\\
 \end{abstract}
 
 \section{Introduction}	
It is well known that the standard perturbation techniques fail to describe the strongly interacting system of quarks and gluons (QGP, similar to a liquid with a small shear viscosity) \cite{CKS} \cite{CGI}. The AdS - CFT correspondence \cite{JM} predicts a universal bound on the ratio $\eta/s$ ($\eta$ - the shear viscosity of the fluid and $s$ - the entropy density), namely $\eta/s~\geq~1/4 \pi$, close to the value obtained from  a hydrodynamic model of the relativistic heavy-ion collisions (RHIC) \cite{LR} \cite {HN} 

 Castorina, Kharzeev and Satz \cite{CKS} showed that at high energies the universal hadronic freeze out temperature $T_{f} \approx 170 MeV$ is a Unruh temperature $T_{U} = a/2 \pi = (\sigma/2 \pi)^{1/2} \approx 170 MeV $, where $a$ is the deceleration of quarks and antiquarks and $\sigma \approx 0.18 GeV^{2}$ is the QCD string tension. In addition, they consider the hadronic Rindler spacetime formed at RHIC is the near horizon approximation of some black hole (BH) geometry. 
 
 Nastase \cite{HN} stated that the fireball observed at RHIC is a dual BH with a temperature proportional to the pion mass and with a value about the experimental ''freeze out'' temperature of $176 ~MeV$. The core of the fireball is the pion field soliton . Moreover, the BH created in the collision has no any singularity at its centre and decay products are all thermally distributed. 

 Luzum and Romatschke\cite{LR} consider dissipative hydrodynamics offer a sensible description of the experimental data for the properties of RHIC. It is a system expanding in a boost invariant fashion (when the hydrodynamical variables are independent on rapidity) along the longitudinal direction (Bjorken expansion). The fluid is thus comoving in the Milne coordinates \cite{KLT} which are most appropriate as the outcome of the collision taking place in the Rindler wedge $t \succ |x|$. In addition, the hydrodynamic variables are independent on the rapidity $y$ and the transverse directions to the expansion. Kajantie et al. conjectured that the 4 - dimensional QCD matter undergoing scale free Bjorken expansion contains a Casimir - type contribution (vacuum energy term) to the holographic stress tensor.
 
 Nakamura and Sang-Jin Sin \cite{NS} have remarked that, since the RHIC fireball is expanding along the collisional axis, we need to understand AdS/CFT in the time dependent regime. In Milne's frame all the fluid points are at rest and, therefore, they share the same proper time since the real fireball produced by RHIC experiments is localized (the central rapidity region playing the basic role).
 
 We mention that the Hawking - Unruh radiation has never been observed in astrophysics so far\cite{CGI}. The thermal hadron spectra in RHIC may thus be the first experimental opportunity to detect such radiation.\\
 Throughout the paper the conventions $G = c = k_{B} = \hbar = 1$ are used.

 \section {Milne geometry from Rindler}
 
 Let us take the Minkowski line element 
 \begin{equation}
 ds^{2} = -dt^{2} + dx^{2} + dx_{\bot}^{2},
 \label{1}
 \end{equation}
 where $dx_{\bot}^{2}$ stands for the two spatial direction orthogonal to $x$ direction.\\
 By means of the coordinate transformation 
 \begin{equation}
 x = (X - \frac{1}{g})~ cosh gT, ~~~t = (X - \frac{1}{g})~ sinh gT ,
 \label {2}
 \end{equation}
 Eq. (1) yields 
 \begin{equation}
 ds^{2} = -(1-gX)^{2} dT^{2} + dX^{2} + dx_{\bot}^{2},
 \label{3}
 \end{equation}
 which represents the well known Rindler metric viewed by a uniformly accelerated observer having a constant rest-system acceleration $g$. The $X = const.$ observers move along the trajectory $x^{2} - t^{2} = (X - 1/g)^{2}$ in Minkowski space. The event horizon $X = 1/g$ corresponds to the two light cones $x = \pm t$. 
 
 The transformation    
 \begin{equation}
 1 - gX = \sqrt{1 - 2g \bar{x}} 
 \label{4}
 \end{equation}
 brings (3) in the form
 \begin{equation}
 ds^{2} = - (1 - 2g \bar{x}) d \bar{t}^{2} + \frac{d \bar{x}^{2}}{1 - 2g \bar{x}} + dx_{\bot}^{2}, 
 \label{5}
 \end{equation}
 where $X \prec 1/g,~ \bar{x} \prec 1/2g$ and $ \bar{t} \equiv T$. 
 
  Let us consider the region where $\bar{x} \succ 1/2g$. In that case $ 1 - 2g \bar{x}$ becomes negative and $\bar{x}$ is timelike and $\bar{t}$ - spacelike. One means we are beyond the horizon $\bar{x} = 1/2g$. Therefore, we replace $\bar{x}$ with $\bar{T}$ and $\bar{t}$ with $\bar{X}$ (the conversion is similar with that encountered in the BH spacetime when the horizon $r = 2m$ is crossed\cite{HC1}). One obtains 
 \begin{equation}
 ds^{2} = - \frac{d \bar{T}^{2}}{2g \bar{T} - 1} + (2g \bar{T} -1) d\bar{X}^{2} + dx_{\bot}^{2}. 
 \label{6} 
 \end{equation}
 
 The above procedure is equivalent to the analytical continuation across the Rindler horizon. We replace now $(\bar{T}, \bar{X})$ coordinates with $(\tau, y)$, according to 
 \begin{equation}
 \sqrt{2g \bar{T} - 1} = g \tau,~~~~\bar{X} = y,
 \label{7}
 \end{equation}
 where $\bar{T} \succ 1/2g$. The spacetime (6) becomes now
 \begin{equation}
 ds^{2} = - d \tau^{2} + g^{2} \tau^{2} dy^{2}  + dx_{\bot}^{2}, 
 \label{8}
 \end{equation}
 which is the Milne metric, well known from cosmology and, more important in our case, from the RHIC (Milne's coordinates are adapted to the Bjorken flow since the velocity vector of the flow is $\partial_{t}$). They are nothing else but the Rindler coordinates in the quadrant $t \succ |x|$, where the stationary observer is located. That observer (beyond the Rindler horizon) detects thermal radiation of temperature $T_{U} = g/2\pi$ as a consequence of the fact that some particle accelerates in the region $|x| \succ t$. 
 
 While the accelerated particle moves in the Rindler wedge, for the stationary observer from the ''hidden'' region the geometry is time dependent (Milne's universe). 
 
 A similar situation we encounter when an electric charge is uniformly accelerated. The retarded radiation emitted by the charge is measured by an observer located in a region inaccesible to the charge but the total energy passing through the surface comoving with the particle is zero \cite{DB} (no electromagnetic energy flux). As Boulware has noticed, the flow of energy comes from the past horizon as if it were another opposite charge in the opposite Rindler wedge. \\
 However, the Unruh radiation is nonvanishing both for the accelerated observer and for the stationary one from the Milne region. As Castorina et al.\cite{CKS} have observed, the stationary observer in the ''hidden'' region measures thermal radiation of Unruh temperature as a consequence of the passage of the accelerated particle. 
  In terms of the Minkowski coordinates, we have
 \begin {equation}
 \tau = \sqrt{t^{2} - x^{2}},~~~~y = \frac{1}{2g} ln \frac{t+x}{t-x}.
 \label{9}
 \end{equation}
 In other words, $\tau$ corresponds to the Minkowski interval (proper time) and $gy$ is the (adimensional) rapidity, with $\tau \succ 0$ and $-\infty \prec y \prec \infty $ ($|y| \approx \infty$ corresponds to the fronts of the expansion fluid\cite{NS} and $y \approx 0$ is the central rapidity region). 
 
 It is a known fact that the Schwarzschild geometry is almost flat near the event horizon $r = 2m$, where the line element appears as 
 \begin{equation}
 ds^{2} = - (\frac{r}{2m} - 1) dt_{S}^{2} + \frac{2m}{r-2m} dr^{2} + 4m^{2} d \Omega^{2} ,
 \label{10}
 \end{equation}
 where $t_{S}$ is the Schwarzschild time , $m$ is the central mass and $d \Omega^{2} = d\theta^{2} + sin^{2} \theta~ d\phi^{2}$ is the metric on the unit two - sphere \cite{HC2}. Using the expression of the surface gravity on the horizon, $\kappa = 1/4m$, eq. (10) gives us
 \begin{equation}
 ds^{2} = - (2 \kappa r - 1) dt_{S}^{2} + \frac{dr^{2}}{2 \kappa r - 1} + r^{2} d \Omega^{2} ,
 \label{11}
 \end{equation}
 with $r \succ 2m = 1/2 \kappa$. The spacetime (11) resembles the Rindler metric (5) when we take $\theta, \phi$ = const. and replace $\kappa$ with the acceleration $g$. This is a well known result. We should only stress the direct connection between the mass and the proper acceleration $g$ \cite{HC3}. If we go further by analogy, we may conclude that for $r\prec2m$ and near the BH horizon, the geometry must be Milne's, which is flat. A time dependent metric inside a BH has been proposed in Ref.8 and seems to be 
 \begin{equation}
 ds^{2} = - d\hat{t}^ {~2} + dz^{2} + \hat{t}^ {~2} d\Omega^{2},
 \label{12}
 \end{equation}
 where $z$ plays the role of the radial coordinate ($-\infty < z < \infty$) and $\hat{t}$ is the temporal coordinate.
    The above geometry is curved (the scalar curvature $R_{\mu}^{\mu} = 4/\hat{t}^ {~2}$) with a singularity on the hypersurface $\hat{t} = 0$. Nevertheless, when $\hat{t} \rightarrow \infty$, the spacetime (12) becomes flat. That could be seen from the expression of the Kretschmann scalar 
 \begin{equation}
 R_{\alpha \beta \mu \nu} R^{\alpha \beta \mu\nu} = \frac{16}{\hat{t}^{~4}},
 \label{13}
 \end{equation}
 computed from the only nonzero component of the Riemann tensor
 \begin{equation}
 R_{\theta \phi \theta \phi} = 2 \hat{t}^{~2} sin^{2} \theta .
 \label{14}
 \end{equation}
 In addition, all the components of the stress tensor inside the BH have the same behaviour : they vanish when $\hat{t} \rightarrow \infty$ 
 
 As it was shown in Ref.8, $\hat{t} \rightarrow \infty$ is equivalent to ''near the horizon'', viewed from the interior of the BH. It is not surprising to get flat spacetime at temporal infinity. We only remind that the ''near horizon'' approximation (10) for the Schwarzschild spacetime leads to a curved metric in spherical coordinates (one has a nonzero component of the Riemann tensor : $R^{\theta \phi}_{\theta \phi} = 1/4m^{2}$).
 The metric becomes flat when $m \rightarrow \infty$ ; that has the same effect as $\hat{t} \rightarrow\infty$ inside the horizon.

  The metric (3) from Ref.8 is curved but when the time tends to infinity the components of the Ricci tensor are vanishing and the geometry becomes Minkowskian. The difference compared to Milne's line element comes from the spherical coordinates used . The fact that the interior of a BH has no any singularity has been recently remarked by Nastase\cite{HN} in his dual BH model. He stressed that the core of the fireball carries informations from inside  the BH, with no singularity at its centre.
  \section{The shear viscosity tensor}
 Our next task is to compute the components of the shear tensor corresponding to the RHIC fireball, in the Milne spacetime (8). The non - null Christoffel symbols we need are
 \begin{equation}
 \Gamma_{y \tau}^{y} = \frac{1}{\tau} ,~~~~\Gamma_{yy}^{\tau} = g^{2} \tau.
 \label{15}
 \end{equation}
 The covariant expression of the shear tensor is given by 
 \begin{equation}
 \sigma_{\alpha \beta} = \frac{1}{2}(h_{\beta}^{\mu} \nabla_{\mu} u_{\alpha}+ h_{\alpha}^{\mu} \nabla_{\mu} u_{\beta})-\frac{1}{3} \Theta h_{\alpha \beta}+ \frac{1}{2}(a_{\alpha} u_{\beta} + a_{\beta} u_{\alpha}),
 \label{16}
 \end{equation}
 where $u_{\alpha} = (-1, 0, 0, 0) $ is the proper velocity of the fluid,  $h_{\alpha \beta} = g_{\alpha \beta} + u_{\alpha} u_{\beta} $ is the induced metric on a hypersurface orthogonal to $u_{\alpha}$ and $a^{\alpha} = u^{\beta} \nabla_{\beta} u^{\alpha} $ (it is vanishing in our situation since the fluid worldlines are geodesics). We have taken a ''comoving'' frame where all the fluid points are at rest and share the same proper time $\tau$ \cite{NS} . In other words, the local rest frame of the fluid is given by $\tau$ and the rapidity $gy$. \\
  The scalar expansion $\Theta$ can be found from 
 \begin{equation}
 \Theta = \nabla_{\alpha} u^{\alpha} ,
 \label{17}
 \end{equation}
 Using (17), the scalar expansion appears as 
 \begin{equation}
 \Theta = \frac{1} {\tau} ,
 \label{18}
 \end{equation}
 with $\dot{\Theta} \equiv u^{\alpha} \nabla_{\alpha} \Theta = -1/\tau^{2}$. The fact that the above $\Theta$ is one half of that one obtained in \cite{HC1} is due to the two time dependent metric coefficients in the BH interior geometry. From the previous equations we obtain
 the nonvanishing components of $\sigma_{\alpha \beta}$ 
 \begin{equation}
 \sigma_{y}^{y} = \frac{2} {3 \tau}, ~~~~\sigma_{\bot} = \frac{-1} {3 \tau},
 \label{19}
 \end{equation}
 where $\sigma_{\bot}$ represents the components on the two transversal directions. We note that $\Theta$ and the shear tensor are divergent on the hypersurface $\tau = 0$. As Beuf et al. \cite{BHJP} have noticed in their study on boost - invariant early time QGP dynamics from AdS - CFT, the initial data must contain a singularity in the bulk of AdS which may signal the presence of a dynamic horizon even at the start of the Bjorken expansion. This strenghtens our view concerning the role played by an event horizon in the formation and evolution of the QGP.
  It is an easy task to check that the Raychaudhury equation 
 \begin{equation}
\dot{\Theta} - \nabla_{\alpha} a^{\alpha}+ 2(\sigma^{2}- \omega^{2})+ \frac{1}{3} \Theta^{2} = - R_{\alpha \beta} u^{\alpha} u^{\beta}
\label{20}
\end{equation}
is observed. The vorticity tensor $\omega_{\alpha \beta}$ and the Ricci tensor $R_{\alpha \beta}$ are vanishing (the Milne spacetime is flat). In (20) we have $2 \sigma^{2} = \sigma_{\alpha \beta} \sigma^{\alpha \beta} = 2/3 \tau^{2}$ and $2 \omega^{2} = \omega_{\alpha \beta} \omega^{\alpha \beta} = 0$.

The components of $\sigma_{\alpha \beta}$ from (19) have the same form with those found in Ref.8 , even though the spacetime is not flat in the latter case. The two angular components $\theta$ and $\phi$ have the same contribution as $x_{\bot}$ from the present case. The Milne geometry is, of course, more appropriate since the Bjorken expansion is more or less one dimensional (along the collisional axis of the heavy ions).

\section{Conclusions}
 We have analysed a direct connection between Rindler's and Milne's flat spacetimes and stressed their basic role played in the Bjorken flow approximation applied to the expanding matter in RHIC. The Milne metric is valid beyond the Rindler event horizon (in the region $t \succ |x|$), where the measurements are performed and it depends upon the proper acceleration $g$ of the uniformly accelerated Rindler observer.

 We have also pointed out the similarities between the late time near horizon BH interior geometry and the flat Milne geometry.
 
 An analogy with the anisotropic fluid from the time dependent BH interior metric is considered and it is shown that the components of the shear tensor are identical in both RHIC and BH interior spacetimes.\\

 \textbf{Acknowledgements}\\
    I would like to thank the Organizers of the International Workshop on Astronomy and  Relativistic Astrophysics, IWARA 09, for inviting me to attend this very stimulating meeting, for their warm hospitality during my stay at Maresias, Brazil, where the paper was presented, and  for financial support.

\end{document}